\documentclass[conference, 10pt, letterpaper]{IEEEtran}
\usepackage[letterpaper, left=1in, right=1in, bottom=1in, top=0.75in]{geometry}
\usepackage[noadjust]{cite}
\usepackage{multirow}
\usepackage{mdwmath}
\usepackage{mdwtab}
\usepackage{rotating}
\usepackage{adjustbox}
\usepackage{graphicx}
\usepackage{amssymb}
\usepackage{balance}
\usepackage[cmex10]{amsmath}
\usepackage{algpseudocode}
\usepackage{amsmath}
\usepackage{bmpsize}
\usepackage{times,multicol,array,float,fancybox,listings}
\usepackage[options]{mcode}
\usepackage{subcaption}

\begin{document}

\title{Utilizing Mobile Nodes for Congestion Control in Wireless Sensor Networks}

\author{\IEEEauthorblockN{Antonia Nicolaou\authorrefmark{1}, Natalie Temene\authorrefmark{1}, Charalampos Sergiou\authorrefmark{1}, Chryssis Georgiou\authorrefmark{1} and Vasos Vassiliou\authorrefmark{1}\authorrefmark{2}}
\IEEEauthorblockA{\authorrefmark{1}Department of Computer Science, University of Cyprus, Nicosia, Cyprus\\
\authorrefmark{2}RISE Research Center on Interactive Media, Smart Systems and Emerging Technologies, Nicosia, Cyprus\\
\{anicol15,n.temene,sergiou,chryssis,vasosv\}@cs.ucy.ac.cy}
}

\maketitle

\begin{abstract}

Congestion control and avoidance in Wireless Sensor Networks (WSNs) is a subject that has attracted a lot of research attention in the last decade. Besides rate and resource control, the utilization of mobile nodes has also been suggested as a way to control congestion. In this work, we present a Mobile Congestion Control (MobileCC) algorithm with two variations, to assist existing congestion control algorithms in facing congestion in WSNs. The first variation employs mobile nodes that create locally-significant alternative paths leading to the sink. The second variation employs mobile nodes that create completely individual (disjoint) paths to the sink. Simulation results show that both variations can significantly contribute to the alleviation of congestion in WSNs.

\end{abstract}

\section{Introduction}
\label{sec:intro}

\noindent Every type of network, inevitably, faces the challenge of traffic congestion. Especially in Wireless Sensor Networks (WSNs) where the resources are limited, congestion control is an important problem that should be tackled effectively since, in the opposite case, it may even ruin the whole functionality of the network.
When congestion occurs, hotspot areas are created and the nodes around these areas face either buffer overflowing situations or channel loading situations that reduce the throughput of the network. To mitigate, or avoid, congestion occurrence, several congestion control algorithms have been proposed \cite{Sergiou_Survey}. This can be done either by controlling the load that the sources produce, or by increasing the resources of the network, usually by creating alternative routing paths by employing nodes that are not in the initial source-to-sink paths.

Another way to mitigate congestion is to increase the capacity of the network by utilizing mobile nodes.
Mobile nodes are nodes that may change their location after their initial deployment and can be applied either by all or a subset of nodes in a WSN. The algorithms developed for using mobile nodes are based on two approaches: using mobile sink(s) or mobile sensor nodes. A mobile sink approach will have a sink node that has the ability to move around the network and request data from the neighbouring nodes that have one or two hops distance from it. The use of mobile sink(s) in the network can mitigate the problem of network disconnection and balance the energy consumption of the network. This approach has three basic patterns: a random mobility model, where a random path is followed by the mobile sink; the predictable/fixed path mobility model where a specific programmed path is followed by the mobile sink in a round robin way; and the controlled mobility model where a controlled or guided path is followed, based on some parameters or events.  The second approach, using mobile nodes, is used to either assist the nodes of the network when help is needed or be a part of it from the beginning. Their presence in the network is very useful in solving congestion or maximizing the lifetime of the network.

Example solutions that use mobile sinks for congestion avoidance are the COngestion avoidance for Sensors with a MObile Sink (CoSMoS) \cite{COSMOS} and the Congestion Avoidance and Energy Efficiency (CAEE) \cite{MobSink} protocols. In the first case the authors suggest a mobile sink, which based on specific techniques like path reconfiguration, load estimation techniques, and transient periods of reduced mobility attempts to avoid congestion by collecting the excessive packets. In the latter case, the CAEE protocol also utilizes the concept of mobile sinks, with the major difference from CoSMoS that the network is divided into clusters called mini-sinks. The cluster head is called a data collector node. The main responsibility of a data collector node is to receive and store the collected data from the sensor field to the mini-sink. The mobile sink periodically visits each mini-sink in the sensor field for data retrieval.

Another effort that employs the concept of mobile nodes for congestion avoidance is the Priority Based Congestion Control Dynamic Clustering (PCCDC) protocol \cite{beulah2015priority}. In this protocol the mobile nodes are also organized dynamically into clusters, where at each round the clusters change and every node maintains an updated neighbor table. The cluster head is responsible of the data collection, the transmission of the data towards the sink and the creation of the TDMA time slot. There are two methods detecting congestion: the intra-cluster method where linear feedback is used for hop-by-hop congestion control and the inter-cluster phase where binary feedback is used for end-to-end congestion control.

The concept of utilizing mobile nodes in the network for the creation of alternative paths to the sink was initially suggested in \cite{koutroullos2011mobile}. In \cite{koutroullos2011mobile}, the Mobile Congestion Control (MobileCC) mechanism is proposed for use in areas that suffer from congestion repeatedly, permanently or for a long duration. The basic idea is that a number of mobile nodes are placed beside the sink, and when congestion occurs, the sink sends the mobile nodes to create hard alternative disjoint paths, consisting only of mobile nodes, to relieve the congestion area from traffic. The initial MobileCC work does not really address the actual mobile node placement strategy. It rather proves (in a before and after fashion) that if mobile nodes are used to create dedicated disjoint paths, it is possible to mitigate the effects of congestion.

In this work we offer a novel solution on how the concept of mobile relay nodes can be used to resolve congestion control. We retain the basic principles of MobileCC and we focus on the last part of the framework, namely the part that reacts to the appearance of congestion and resolves the problem by efficiently and effectively relocating mobile nodes.
The Alternative Path Creation mechanism starts when existing congestion control algorithms fail and it consists of two variations: a dynamic node placement algorithm that solves the problem locally and a direct node placement algorithm that creates a new direct path to the sink, which consists only of mobile nodes.

The rest of this paper is organized as follows. In Section~II the MobileCC mechanism is explained. Section~III offers details on the mobile node placement algorithms we introduce and in Section~IV an initial evaluation of the proposed algorithms is presented. Finally, the conclusions of this work
are discussed in Section~V.

\section {MobileCC Framework}
\label{sec:framework}

\noindent We consider a network that consists of randomly deployed static nodes and a small set of mobile nodes residing next to the sink. We also assume the following:

\begin{itemize}
\item All nodes, both static and mobile, are identical in terms of computation power, communication capabilities, sensing and transmission range etc. with the exception that mobile nodes can move.
\item We employ a simple MAC protocol, like CSMA/CA.
\item All nodes are aware of their location in relation with the location of the sink.
\end{itemize}

The primary objective of this work is to utilize the extra resources (mobiles nodes) efficiently and effectively in order to resolve congestion in the network and if possible to improve its performance in terms of delay, energy efficiency and throughput. The problem has two aspects. The first aspect is the placement of the mobile nodes in such a way as to create a disjoint path made up entirely of mobile nodes to the sink, while in the other case, the mobile node creates a local disjoint path that connects with the original routes.

\noindent MobileCC is composed of the following mechanisms:
\begin{itemize}
  \item Congestion Detection Mechanism
  \item Congested Node Selection Mechanism
  \item Congestion Notification Mechanism
  \item Alternative Path Creation Mechanism Using Mobile Nodes
    \begin{itemize}
    \item Calculation of Extra Resources
    \item Calculation of Optimum Position of Extra Nodes
    \item Establishment of Alternative Path
    \end{itemize}
\end{itemize}

Concerning the three first mechanisms, a lot of work has already been done so far \cite{Sergiou_Survey}. In this work we choose to employ the simple, yet efficient, mechanisms already introduced in algorithm DAlPaS~\cite{Dalpas_Jour}. DAlPaS employs a dynamic way to control topology without adding any extra load to the network. To do this, it uses data from the neighbor tables that have been created during the setup phase and especially the level that a node resides (i.e. the number of hops away from the sink). Thus, every node that is going to transmit data searches in its neighbor table and finds the node with the lowest level (which is closer to the sink) and transmits its data through this node. If multiple nodes are available, the choice is made using a tie-break mechanism based on the smallest node ID. As a result, a dynamic spanning tree is being created and each node transmits its data through the shortest path.

\begin{figure}[h!]
	\centering
	\includegraphics[scale=0.4]{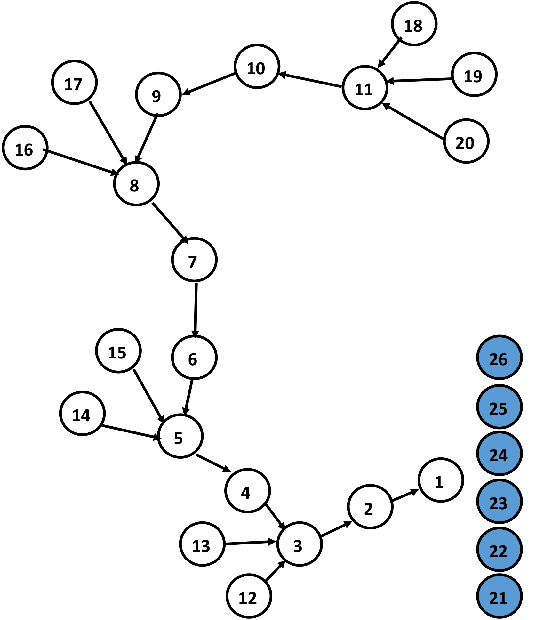}
	\caption{Example Network Topology}\label{fig:Init}
	\label{fig:exinitialtopology}
\end{figure}

For a better explanation of the different mechanisms, let us take as an example the network shown in Figure~\ref{fig:exinitialtopology}. This is an instance of a network in which a relatively small number of nodes remain active and there is essentially only a single path leading to the sink (node~1). It is not difficult to recognize that certain nodes in this topology (nodes 3, 5, 8, and 11) are possible congestion hotspots and may need assistance from the sink. Mobile nodes are shown in blue color adjacent to the sink (node~1).

\section {Alternative Path Creation Mechanism Using Mobile Nodes}

\noindent When a node detects congestion, it sends a Congestion Message (CM) to the sink. This message contains all the information needed from the sink to act for mitigating the congestion that has appeared in the network. This information includes: the congested node's NodeID, the number of packets received and forwarded per sample time period, the timestamp of when the congestion detection has started, and its neighbor table information. From the neighbor table, the information included is: the neighbor nodes' NodeIDs, hop number, number of packets received and availability flag. When the network sink receives the CM message it calculates the position that it would be ``clever" to place a mobile node in order to provide alternative paths to the sink for mitigating congestion. When the position calculation is ready, the sink sends a message to the mobile nodes specifying the target location, the sender nodes' NodeIDs, and the next hop NodeID. When mobile nodes receive the information from the sink, they switch off their radio, move towards the target positions, and turn their radio back ON. Using this OFF/ON tactic, they are not detectable by the existing network nodes while they move, nor do they create any interference while travelling towards the target locations. Finally, when the mobile node reaches its destination it will establish a connection with the nodes to be served.

In this section we describe two algorithms, which can be used for determining the number and position of mobile nodes.\smallskip

\subsection{Dynamic Node Placement (locally-significant paths)}
\label{sec:Dynamic}
\noindent Initially, we propose the Dynamic Node Placement algorithm, referred to as Dynamic MobileCC. This algorithm places a mobile node in such a position, so as to receive traffic from the nodes that transmit data to the congested node. This mobile node can forward the packets directly to the sink, if the sink is in its transmission range, or it can serve as a relay node, forwarding the received packets to other upstream nodes.\smallskip

\noindent {\bf\em High level idea.}
Initially, the Dynamic MobileCC algorithm calculates the average number of packets per time unit that the congested node receives and cannot forward due to lack of buffer space.
Then, it discovers the nodes that transmit their packets to the congested nodes and calculates the best position that the mobile node(s) should move to in order to receive data from a number of them. Ideally, the best position is the position where the minimum number of nodes can divert their traffic through the mobile node(s), whereas at the same time their total sending rate should be equally or more than the amount of the excess traffic of the congestion node.

Its operation is based on the following functions:
\begin{itemize}
\item Identification of congested and ``congesting'' nodes
\item Calculation of extra resources
\item Calculation of the optimum position that the mobile node should be placed
\end{itemize}

We now provide a detailed description of these functions below.\smallskip

\subsubsection{Identification of congested and ``congesting'' nodes}

The first step in this algorithm is the identification of the node that is congested and the nodes that congest this node. This operation is normally performed by existing congestion detection algorithms. This information, along with the position of these nodes is communicated to the sink. For example, a simple and efficient way is the routing table that is being used in
algorithm DAlPaS~\cite{Dalpas_Jour}.\smallskip

\subsubsection{Calculation of extra resources}

Then, the average number of packets per time unit (e.g., seconds) that the congested node receives and cannot forward, is calculated. Based on this, the algorithm calculates the $Additional\ Resources$ that are required to accommodate the excess traffic. In particular, for a {\it congested node} $i$, the $Additional\ Resources$ rate $A(i)$ is calculated by the equation:

\begin{equation}
\label{eq:additional}
A(i) = \frac{Recv(i)- Tran(i)}{t - t_0}
\end{equation}
\noindent
where, \\
$Recv(i)$ is the number of packets that $i$ has received from its neighbors, \\
$Tran(i)$ is the number of packets that $i$ has transmitted,\\
$t$ is the current time, and \\
$t_0$ is the time that $i$ started transmitting packets.\medskip

Based on this equation, the mobile nodes that will move close to the congested hotspot, should be able to receive and forward the excess traffic load that cannot be forwarded by the congested node. Thus, the congested node will receive just the traffic it can accommodate and congestion will be alleviated.\smallskip

\subsubsection{Calculation of the position that the mobile node should move to}

The algorithm checks whether there is a single node, which if it stops transmitting towards the congested node, congestion will be alleviated. If there is such a node then the single point where the mobile node should move is calculated. Otherwise, if there are more than one nodes, then for each of these nodes a specific point is calculated.  The objective of this algorithm is to minimize the number of nodes that will transmit data through the mobile nodes, but their total sending rate should be equal to the amount of traffic that the congested node is not able to forward, hence eliminating congestion. Furthermore, the mobile nodes should be placed in a position where at least a non-congested node should exist in their transmission range, so as to forward the data they receive, to the sink.

\begin{figure*}[t]
  \centering
  \begin{subfigure}{.6\columnwidth}
    \centering
    \includegraphics[width=\linewidth]{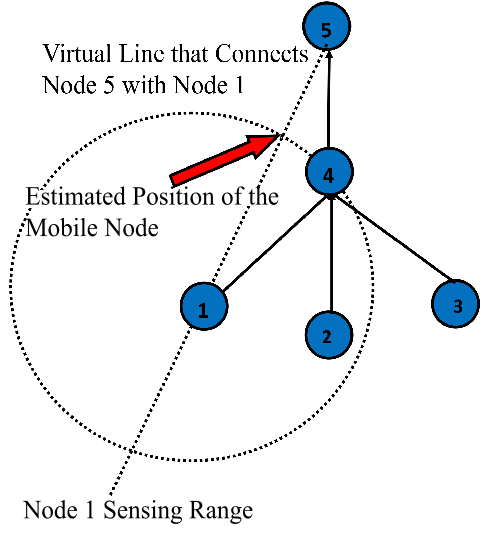}
    \caption{Single Node}
  \end{subfigure}%
  \hfill
  \begin{subfigure}{.7\columnwidth}
    \centering
    \includegraphics[width=\linewidth]{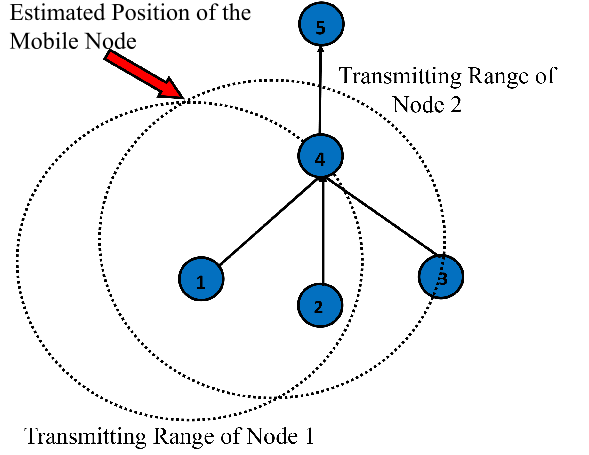}
    \caption{Multiple Node}
  \end{subfigure}%
  \hfill
  \begin{subfigure}{.7\columnwidth}
    \centering
    \includegraphics[width=\linewidth]{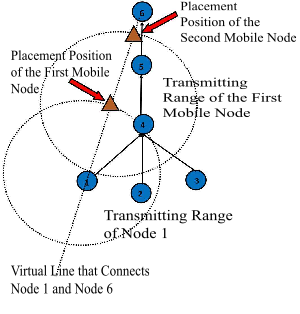}
    \caption{Direct Path}
  \end{subfigure}
  \caption{Node Placement Positions}
  \label{fig:fig1}
\end{figure*}

The calculation of this specific point is performed as follows:

Initially, the intersection points between the circle that is created by the radius of the transmitting range of the congested node and the straight line that connects the sink with this node, is calculated. Between these two points, the point which is closer to the destination node in comparison to the point which is closer to the mobile node, is chosen (Fig \ref{fig:fig1}a).

Let's consider as $(X_{k}, Y_{k})$ the coordinates of the node that is going to be served by the mobile node and $(X_{sink}, Y_{sink})$, the coordinates of the sink. In case that the coordinate $X$ of this node, or $Y$ respectively, is the same as of sink's, i.e. if $X_{k}=X_{sink}$, then the intersection point will be $(X_{k}, Y_{k} + node's\ Tx\ range)$ if $Y_{k}< Y_{sink}$ and $(X_{k}, Y_{k} -node's\ Tx\ range)$ if $Y_{k}>Y_{sink}$. If $Y_{k}= Y_{sink}$, then the intersection point will be $(X_{k}- node's\ Tx\ range,Y_{k})$ if $X_{k}> X_{sink}$ and  $(X_{k}+ node's\ Tx\ range,Y_{k})$ if $X_{k}< X_{sink}$.

For each node that has a sending rate greater than the $Additional\ Resources$ rate, the position of the mobile node is calculated and the algorithm checks whether there is a node closer to the sink which is not congested, so as to transmit the data that it receives.

\subsubsection{ Finding the position where mobile nodes should move in order to serve more than one nodes}
If there is not any available relocation position of the mobile node suitable to serve just one node, then for a number of nodes equal to $n$, where $n$ is a number between $2$ and $6$ according to \cite{DBLP:journals/tcom/TakagiK84} and ~\cite{DBLP:journals/tcom/HouL86}, the following procedure is followed:

Initially, the algorithm described in \cite{knuth2005art} is employed. This algorithm identifies the subset of the nodes that transmit their data to the congested node. Only the subsets that have a total sending rate greater than the $Additional\ Resources$ rate of the congested node are used for calculations.
For each of these subsets, the algorithm finds the common point in the transmission range of the nodes, which is closer to the sink. To achieve this, the algorithm considers for each pair of these nodes, the cross-section of their transmitting ranges. Then, it checks whether this cross-section point is within the transmitting range of the rest of the nodes, besides the pair under reference. If, for a subset of nodes, more than one appropriate point is calculated, then the point which is closer to the sink is chosen.
This is illustrated in Fig. \ref{fig:fig1}b.

An example of placing a mobile node for more than one nodes based on Fig.~\ref{fig:fig1}b is described below. Node 4 is congested because it receives more packets from nodes 1,2,3 that it can handle. At first, the algorithm checks for subsets with size 2 and creates the subsets $\{1,2\},\{1,3\},\{2,3\}$. We assume that only the total sending rate of nodes 1 and 2 is greater than the $Additional\ Resources$ rate (eq. (\ref{eq:additional})) and that the position of each node in the network is as follows: ($0,0$) for node 1, ($2,0$) for node 2 and ($2,46$) for node 5 which is the sink. We also assume that the transmitting range is $2.5$. The cross-section points calculated from the transmitting range of node 1 and node 2 are $\{2.291,1\},\{-2.291,1\}$. The point selected for the position of the mobile nodes is the one that is closer to the sink. If there were no subsets of size 2 that would have a total sending rate more than the additional resources rate, then the algorithm would check for subsets of size 3, and hence take subset $\{1,2,3\}$.

The procedure halts when at least a common subset of nodes $n$ is found, for $n\in [2,6]$. If there is more than one subsets of size $n$, and more than one common point, then a mobile node is chosen to move to the common point that is closer to the sink. Thus, the algorithm makes sure that, from the smallest subsets ($n=2$) to the largest subset ($n=6$), the subset that is being served by the mobile node is the smallest. This attribute secures the validity of the first limitation of this algorithm, that the least number of nodes should change destination node.\smallskip

\subsection{Direct Node Placement Algorithm}
\label{sec:Direct}
\noindent The Direct Node Placement algorithm is the second variation of the MobileCC mechanism; we refer to it as Direct MobileCC. Similarly to the Dynamic MobileCC  algorithm, it does not replace any existing topology, congestion control or routing algorithms, but runs alongside them.
The difference from Dynamic MobileCC is that it creates a completely new and direct (disjoint) alternative path of mobile nodes towards the sink.
In this way, it is faster in establishing a connection to the sink. As our experimental evaluation shows (c.f., Section~\ref{sec:eval}),
this helps to reduce the number of dropped packages, trading however, use of resources (and hence, energy).\smallskip

\noindent {\em\bf High level idea.}
Initially, the Direct MobileCC algorithm runs the Dynamic MobileCC algorithm to calculate the position of the first mobile node that will be placed in the network. Then, it creates the direct line starting from the first placed mobile node and ending to the sink. On this line it places additional mobile nodes until one of them is in the range of the sink and can forward packets directly to it.

Its operation is based on the following functions:

\begin{itemize}
\item Calculation of the position of placement of the first mobile node using the Dynamic MobileCC algorithm.
\item Creation of a path consisting of mobile nodes, starting from the first mobile node, that was placed from the previous function and ending at the sink.
\end{itemize}

We proceed to describe the second function (since the first is identical to the Dynamic algorithm's function):

If the mobile node placed in the previous function is at the range of the sink, it transmits the received data directly to the sink and the process terminates. If more mobile nodes are needed in order to create a disjoint path to reach the sink, the algorithm calculates the placement position of the next mobile node that should be within the transmitting range of the first node.
The calculation of this specific point is performed as follows:
The intersection points of the virtual circles created by the transmitting range of the initially placed mobile node and the virtual straight line between this node to the sink, is calculated. Between these two points the point which is closer to the sink is kept. This is illustrated in Figure~\ref{fig:fig1}c. The process continues as long as the sink is not reached.

\section{Evaluation}
\label{sec:eval}

\subsection{Evaluation Setup and Results}
\label{subsec:setup}
\noindent We have implemented the two variations, Direct and Dynamic, within the Contiki OS~\cite{contikiwebsite}, an open source operating system for implementing networked, resource-constrained systems, mainly focusing on low-power wireless Internet of Things devices. The evaluation has been performed in the COOJA simulator, a dedicated simulator for Contiki OS nodes. The simulator parameters are presented in Table~\ref{tab:params}. The network topology is the one already used in explaining the algorithms.

\begin{table}[h]
	\begin{center}
		\caption{Simulation Parameters}
		\label{tab:params}
		\begin{tabular}{ll}
			\hline
			Simulator/OS & COOJA/Contiki 3.0 \\
			Protocol & Contiki Multihop/Rime \\
			MAC	& ContikiMAC/CSMA \\
			Simulation Time  & 600 \\
			Repetitions of simulation  & 50 \\
			Emulated Mote & Tmote sky \\
			Number of Nodes (Sink/Fixed/Mobile) & 1/19/6 \\
			Transmission Range (m) & 25 \\
			Max Data Rate (kbps) & 250 \\
			Queue Length & 8 Pkts\\
			\hline
		\end{tabular}
	\end{center}
\end{table}

\subsection{Resulting Topology}

\noindent Initially, we employed 26 Tmote Sky nodes (1 sink, 19 fixed and 6 mobiles nodes) according to the topology of Figure~\ref{fig:exinitialtopology}.

In this scenario there are 9 source nodes (nodes 12-20), and 6 mobile nodes (nodes 21-26). The mobile nodes are placed near the sink in a sleep mode until needed. Two nodes (3,8) become congested due to their placement in the network.

Since there are not, to the best of our knowledge, other algorithms in the literature directly comparable with the algorithms presented in this paper, we have chosen, for comparison purposes to employ DAlPaS \cite{Dalpas_Jour}, an algorithm that creates alternative paths, in case of congestion and routes through them the excess traffic. As we described before, when congestion occurs in the network and there are no further available resources in the network, DAlPaS comes to stall state.  This is the point where our two proposed algorithms are going to run (in fact, are needed to run).

Next we present the derived topologies after the execution of the experiment.

\begin{figure}[h]
	\centering
	\includegraphics [scale=0.4]{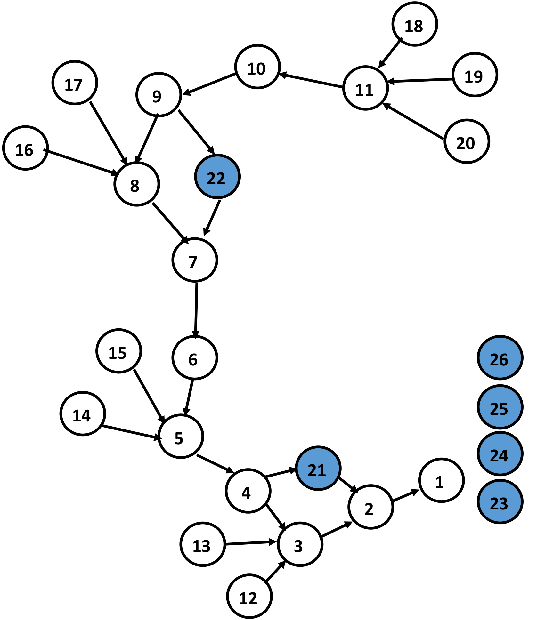}
	\caption{Dynamic MobileCC Execution of the Example}
	\label{fig:dynamictopology}
\end{figure}

In Figure~\ref{fig:dynamictopology} we present the topology, after the sink calls the Dynamic MobileCC algorithm. In this scenario, two mobile nodes are employed, one for each occurrence of congestion.

\begin{figure}[h]
	\centering
	\includegraphics [scale=0.4]{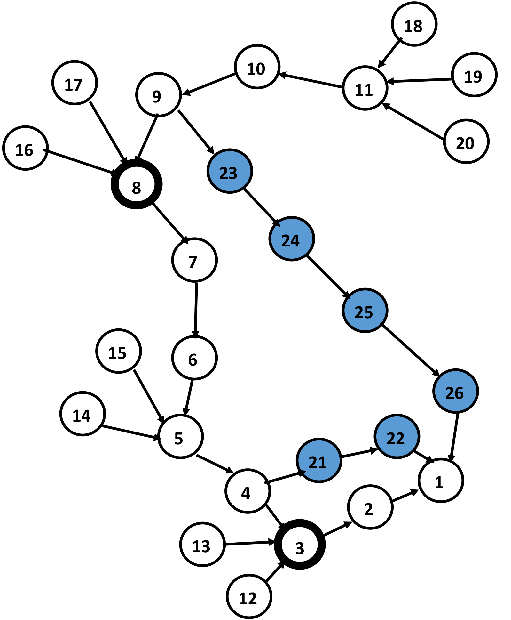}
	\caption{Direct Path MobileCC Execution of the Example}
	\label{fig:directtopology}\vspace{-1em}
\end{figure}

In Figure~\ref{fig:directtopology} we present the topology after the sink calls the Direct MobileCC algorithm. In this case two alternative mobile node paths are created. The first path consists of two mobile nodes and the other one consists of four mobile nodes. The different number of mobile nodes used for each path is related to the distance of the congested node from the sink. Cumulatively, this algorithm employs six mobile nodes for the creation of two disjoint paths to solve the congestion problem.

This simple experiment demonstrates that both Dynamic and Direct MobileCC algorithms can solve the problem locally. Both algorithms must employ at least one mobile node for each congestion occurrence in the network. This example indicates that Direct needs more mobile nodes than Dynamic, which is expected as the former implements a full path of mobile nodes from the congested point to the sink.

\subsection{Numerical Results}

\noindent For the previous mentioned example we also present some basic numerical results.

\begin{figure}[h]
	\centering
	\includegraphics[width=0.45\textwidth]{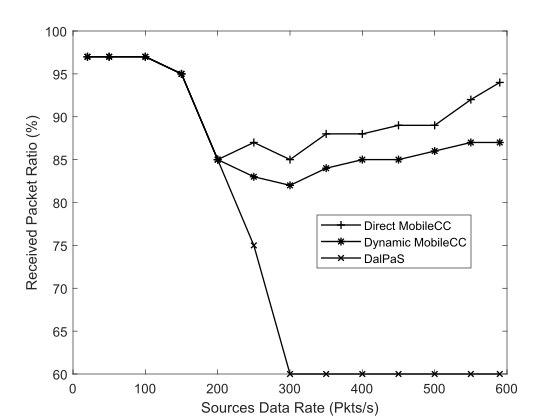}
	\caption{Average Throughput}
	\label{fig:received}\vspace{-1em}
\end{figure}

Figure~\ref{fig:received} shows the ratio of packets received (over all the packets generated by the sources in the course of the simulation) versus the load of the network. We observe that as the network load  (i.e., sources' data rate) increases there is a point, when the data rate reaches 150 pkts/sec (i.e., $150 \cdot 128$ bytes/sec = 2400 bits/sec) at which the DAlPaS algorithm fails to find alternative paths in the existing topology and the network enters deep congestion where essentially no packet reaches the sink (when the rate is 300 pkts/sec). It is shown that both Direct MobileCC and Dynamic MobileCC can relieve the network from the congestion occurrence and maintain at a high level the packet transmissions. The Direct MobileCC algorithm manages a received packet ratio of just 94\% less than the original 97\%. It is worth mentioning that Direct MobileCC delivers more received packets than Dynamic MobileCC. This was expected, since Direct MobileCC creates new disjoint paths of mobile nodes to the sink. In this case, any new appearance of congestion hotspot, through this path is avoided. On the other hand, the Dynamic MobileCC algorithm places just the required number of nodes in specific points of the network, targeting in creating new paths, routing traffic through the nodes that they were not initially accessible. In such a case, congestion may re-appear, especially in cases where some of these nodes are already in use by other flows.

\begin{figure}[h]
    \centering
    \includegraphics[width=0.45\textwidth]{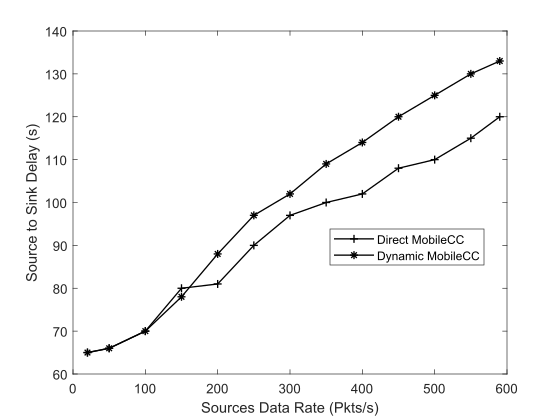}
    \caption{Source to Sink Delay}
    \label{fig:Dropped}\vspace{-1em}
\end{figure}

In Figure~\ref{fig:Dropped} we present the total source to sink delay in the network.
In this plot we notice that both algorithms have a total source to sink delay that increases as a function of the source data rate. This is normal due to the fact that collisions exist in the network, and until the network stabilizes with the help of the mobile nodes, many packets are either resend or sometimes even dropped. As mentioned before, Dynamic MobileCC places only mobile nodes in positions where paths are created from existing nodes in the network so the delay is higher in comparison to Direct MobileCC that creates a new path with mobile nodes.

\begin{figure}[h]
    \centering
    \includegraphics[width=0.45\textwidth]{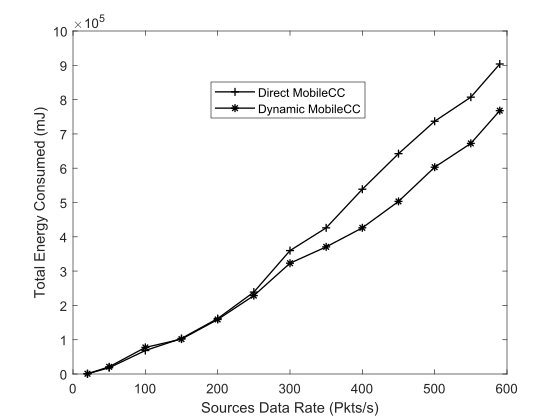}
    \caption{Total Energy Consumed}
    \label{fig:nenergy} 
\end{figure}

In Figure~\ref{fig:nenergy} we present the total energy consumed, measured in $mJ$, during the operation of the network.
To measure the energy consumption of the network, we calculated the energy ($energy_i$) consumed by each node $i$ with the equation bellow \cite{DBLP:journals/adhoc/RazaWV13}:\vspace{-.7em}

\begin{equation*}
\begin{aligned}
energy_i= (transmit * 19.5mA + listen * 21.8mA  + \\
CPU * 1.8mA + LPM * 0.0545mA) * 3V /4096*8,
 \end{aligned}
\end{equation*}
where $trasmit$ is the total time of the radio transmitting, $listen$ is the total time of the radio listening, $CPU$ is the total time of the CPU being active, and $LPM$ is the total time of the CPU being in low power mode.
Then,\vspace{-1em}

\begin{equation*}
\begin{aligned}
Total Energy  = \sum_{i=1}^{n } energy_i \ .
\end{aligned}
\end{equation*}
\noindent

In this plot we observe that both Direct MobileCC and Dynamic MobileCC have a stable increment based on the total packets injected in the network. In comparison, Direct MobileCC has higher energy consumption than Dynamic MobileCC. That was expected, as Direct MobileCC injects more mobile nodes in the network by creating a new alternative path consisting of only mobile nodes.

\subsection{Random Topology}

\begin{figure}[h]
	\centering
	\includegraphics [scale=0.65]{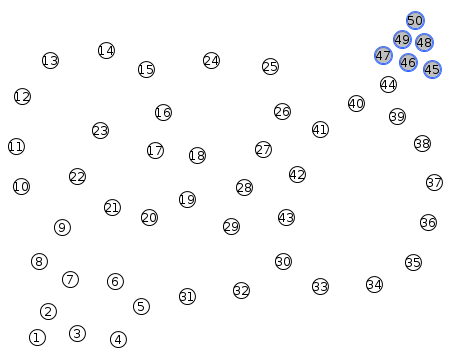}
	\caption{A Uniform Topology of 50 nodes}
	\label{fig:mesh1}
\end{figure}

\noindent To validate the previous results a more realistic scenario has been developed. In this scenario a 10x10 uniform topology has been employed with 50 nodes (Figure~\ref{fig:mesh1}). The source nodes have been selected with a probabilistic function and placed in the lower left area of the network. The sink has been placed in the upper right area, to force the packets flows to converge or intersect. We again use algorithm DAlPaS to indicate the baseline operation.

\begin{figure}[h]
	\centering
	\includegraphics[width=0.45\textwidth]{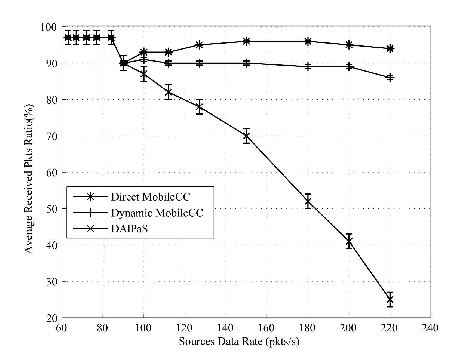}
	\caption{Average Throughput}
	\label{fig:123} 
\end{figure}

In Figure~\ref{fig:123} we present the average throughput.
We notice that after the first congestion occurrence event, DAlPaS is unable to create any other alternative paths to the sink and relieve the problem. As a result, as the traffic in the network increases, the throughput is approaching zero since the network becomes heavily congested.

On the other hand, when the two extension algorithms, Direct MobileCC and Dynamic MobileCC are called when DAlPaS fails, the network recovers and it is able to continue its operation.

Between the two algorithms, we notice that Direct MobileCC  is able to constantly deliver higher number of packets to the sink. 

Furthermore, the throughput of the network, when Direct MobileCC is employed, stabilizes sooner than when Dynamic MobileCC is employed. This happens,  as discussed above, because Direct MobileCC creates a disjoint alternative path to the sink, away from the neighbour nodes of the congested node. On the other hand, Dynamic MobileCC only uses the lowest possible number of mobile nodes for each congestion event in the network and needs more time to create an alternative path. Placing a mobile node may solve a specific congestion problem but it can also create another one at a later stage, an issue that Direct MobileCC pre-handles by design.

\begin{figure}[h]
	\centering
	\includegraphics[width=0.45\textwidth]{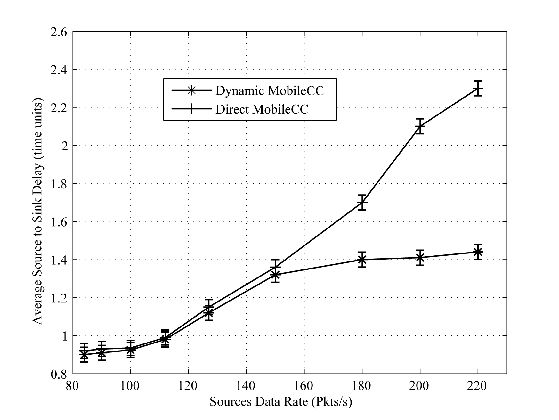}
	\caption{Average Source to Sink delay}
	\label{fig:delay}
\end{figure}

In Figure~\ref{fig:delay} we present the average source to sink delay.
We observe that the average time needed for a packet to be transmitted from a source node to the sink, when  Dynamic MobileCC is employed is higher than the time needed when the Direct MobileCC is employed.

This is normal, since the alternative path created in Direct MobileCC is a direct, disjoint path and the mobile nodes are placed in the transmission range of each of the previously placed mobile nodes. As a result, Direct MobileCC employs less hops in its alternative paths.

\begin{figure}[h]
	\centering
	\includegraphics[width=0.45\textwidth]{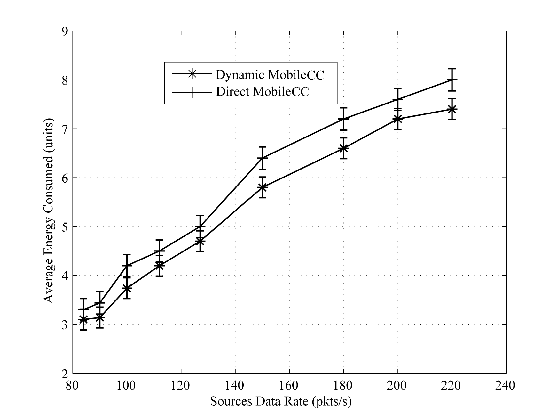}
	\caption{Average Total Energy Consumed}
	\label{fig:Energy}
\end{figure}

In Figure~\ref{fig:Energy} we present the average total energy consumed during the operation of the network.

Also in this case, we notice that the results are consistent with the previous discussion. Direct MobileCC presents higher energy consumption than Dynamic MobileCC, since more mobile nodes are involved in the transmission of packets from source to sink.

\begin{figure}[h]
	\centering
	\includegraphics[width=0.45\textwidth]{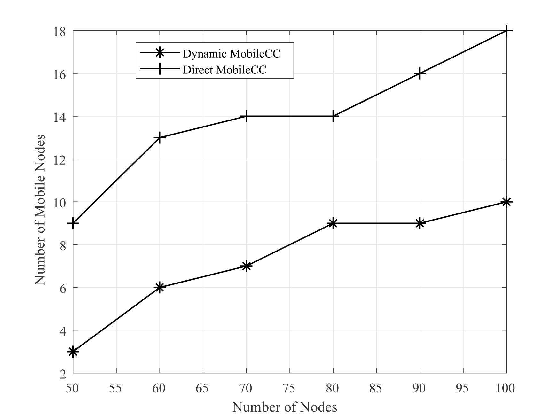}
	\caption{Number of Mobile Nodes Used}
	\label{fig:MNUsed}
\end{figure}

Finally, we present the number of the mobile nodes used by the two algorithms.
In total there were 6 scenarios, with different numbers of nodes in the network. These scenarios consisted of 50, 60, 70, 80, 90 and 100 nodes. We ran each scenario 20 times and the results were taken in average.
In Figure~\ref{fig:MNUsed} we notice that Direct MobileCC employs almost double the number of mobile nodes in comparison with Dynamic MobileCC to resolve the same congestion problem.

\section {Conclusion and Future work}
\noindent In this paper we examined the concept of using mobile nodes in the network to alleviate congestion in WSNs.
We present a mechanism with two variations, that gets initiated when existing congestion control algorithms fail. The mechanism employs mobile nodes to either create disjoint paths of mobile nodes and route the excess traffic directly to the sink (Direct MobileCC), or to place a mobile node in such a position to create alternative path by bridging two disjointed areas in the network, and repeat the process if necessary (Dynamic MobileCC).
Simulation results demonstrate that both variations can alleviate congestion. In doing so, Direct MobileCC demonstrates better average source to sink delay and reduced packet drop, in the
expense of mobile nodes used (almost double) and energy consumed, when compared to Dynamic MobileCC.
In this work we have considered one instance of using alternative paths for alleviating congestion. Future work will extend our solution to consider longer periods of congestion and will include the notion of mobile node re-use and a thorough consideration of the energy cost of each algorithm in such periods.

\section*{Acknowledgment}
\noindent This work is partly supported by the EU's Horizon 2020 programme under grant agreement Nº739578 and the government of the Republic of Cyprus through the Directorate General for European Programmes, Coordination and Development.

\bibliographystyle{IEEEtran}


\end{document}